\documentstyle[emlines]{article}
\begin{document}
\font\twmb=msbm10 scaled \magstep1
\font\temb=msbm10
\font\eimb=msbm8
\def\a{{\cal A}}
\def\j{{\cal J}}
\def\z{\hbox{\temb Z}}
\def\Z{\hbox{\lamb Z}}
\def\ze{\hbox{\eimb Z}}
\def\h{\hbox{\temb H}}
\def\c{\hbox{\temb C}}
\def\r{\hbox{\temb R}}
\def\m{{\cal M}}
\def\n{{\cal N}}
\def\p{{\cal P}}
\def\Re{\hbox{\ Re}}
\def\k#1{\hbox{Ker}\,#1}
\def\v{\hbox{\frak V}}
\def\T{\hbox{\rm Tr}}
\def\tr{\T}
\begin{titlepage}
\begin{center}
\huge \bf
Noncommutative Geometry \\
and The Ising Model
\end{center}

\vspace{1cm}

\begin{center}
\Large \sl
Andrzej Sitarz
\footnote{e-mail: ufsitarz@plkrcy11.bitnet} \\
 Department of Field Theory \\
 Institute of Physics, Jagiellonian University \\
 Reymonta 4, 30-059 Krak\'ow, Poland
\end{center}

\vfill

\begin{abstract}
The main aim of this work is to present the
interpretation of the Ising type models
as a kind of field theory in the framework of
noncommutative geometry.
We present the method and construct sample
models of field theory on discrete
spaces using the introduced tools of discrete
geometry.
We write the action for few models,
then we compare them with various models
of statistical physics. We construct also
the gauge theory with a discrete gauge group.
\end{abstract}
\vfill
\begin{flushleft}
\Large \sc TPJU 18/92
\end{flushleft}
\vfill
\end{titlepage}

\section{INTRODUCTION}

The noncommutative geometry could be considered as a
set of mathematical tools, which, applied to
theoretical physics, can significantly improve and
enlarge the possibilities of model-building in the
field theory [1-4]. These methods allow us to apply
the instruments of differential geometry not only
for the manifolds but also for many non-standard objects
like the discrete spaces and quantum spaces.
For instance, it appears that the Standard Model of
electroweak interactions can be properly described by
the product of continuous and discrete geometry
[2-6], thus suggesting the significance of the
role of the discrete spaces in physics.

In our earlier work \cite{JA} we constructed the necessary
tools to build sample models in the framework of discrete
geometry and we used them to construct gauge theories.
Now, we want to turn our attention to the already existing
class of physical models, which are also situated
on discrete spaces, the most known example being the Ising Model
\cite{HUANG}. The problem, which we would like to consider
in this work is whether such models could be reformulated as
a field theory constructed along the rules of noncommutative
geometry. Our attempt is to show the general way of such
construction and to illustrate it with simple examples.
We also include a brief account of the differential calculus
and the metric properties of discrete spaces.

\section{DIFFERENTIAL CALCULUS}
\def\f{{\cal F}}
This section is devoted to the brief review of the
differential calculus on discrete spaces. We state
here only the most important facts and results, details
could be found in our earlier work \cite{JA}.

Let $G$ be a finite group and ${\a}$ be the algebra of functions on $G$,
which are valued in a field $\f$. The natural choice for $\f$
is the field of complex numbers $\c$, however, one may as well
consider other possibilities.
We denote the group multiplication by $\odot$ and the size
of the group by $N_G$. The right and left multiplications
on $G$ induce natural automorphisms of $\a$, $R_g$ and $L_g$,
respectively,
\begin{equation}
\left( R_h f \right)(g) \; = \; f(g \odot h),
\label{a0} \end{equation}
with a similar definition for $L_g$.

Now we shall construct the extension of $\a$ into
a graded differential algebra. First we introduce the
space of one-forms as a free left-module over $\a$, which is
generated by the elements $\chi^g$, $g \in G'$,
where by $G'$ we denote $G \setminus \{e\}$.
Then we define the external derivative $d$ on the zero-forms
(elements of $\a$) in the following way:
\begin{equation}
d a \; = \; \sum_g \left( a - R_g(a) \right) \chi^g.
\label{a1}\end{equation}
The external derivative is nilpotent and obeys the Leibniz
rule provided that the module of one-forms
has simultaneously a structure of a right-module, as defined for its
generators:
\begin{equation}
\chi^g a \; = \; R_g (a) \chi^g, \;\;\; a \in \a,g \in G',
\label{a3} \end{equation}
and the action of $d$ on the generators $\chi^g$ is as follows:
\begin{equation}
d \chi^g  \; = \; - \sum_{h,k} C_{hk}^g \chi^h \otimes \chi^k,
\;\;\;\; g \in G',
\label{a4} \end{equation}
where the constants $C_{hk}^g$ are the structure constants, obtained
from the relations:
\begin{equation}
(1 - R_i)(1 - R_j) \; = \;
\sum_k C_{ij}^k (1 - R_k).
\label{a5} \end{equation}
In the case of the discrete group $G$ their form is rather simple:
\begin{equation}
C_{hk}^g \; = \; \delta^g_h + \delta^g_k - \delta^g_{(k \odot h)}.
\label{a6} \end{equation}
As already seen in the formula (\ref{a4}) the higher-order forms
are the tensor products of the lower-order ones. Then, the
external derivative acts on them in accordance with the graded
Leibniz rule:
\begin{equation}
d ( v \otimes w ) \; = \; (dv) \otimes w + (-1)^{\hbox{\tiny deg\ }v}
v \otimes (dw).
\label{a7} \end{equation}

The conjugation in the algebra of forms is taken to be
the internal conjugation within the algebra $\a$ for
the zero-forms. For higher-order forms it is
sufficient to define this operation
for the generating one-forms:
\begin{equation}
(\chi^g)^\star \; = \; - \chi^{(g^{-1})}.
\label{a8}
\end{equation}

All these rules give us the structure of the infinite-dimensional
differential algebra over the algebra $\a$. We shall use them as
tools to define simple examples of field theories.

\section{METRIC ON DISCRETE SPACES}

In this section we shall briefly outline the general scheme
of the construction and the properties of the metric. We
shall give the definitions as well as the intuitive picture.

We define the metric on the module of one-forms, as
a middle-linear, \mbox{$\a$-valued}
functional $\eta$:
\begin{eqnarray}
\eta(a \omega_1 , \omega_2 b) \; &=& \; a \eta(\omega_1,\omega_2) b,
\label{me1} \\
\eta(\omega_1 a, \omega_2 ) \; &=& \;  \eta(\omega_1, a \omega_2),
\label{me2}
\end{eqnarray}
This definition is suitable only for the considered case and
it has to be modified for other algebras.
 Both conditions are straightforward generalizations of linearity
requirements for the bimodules.
In the case of discrete geometry, with the module of one-forms
generated by the forms $\chi^g$, the metric is
completely determined by its values on the generators,
$\eta^{gh} = \eta(\chi^g,\chi^h)$. Now, because of (\ref{me2})
and the rules of the differential calculus (\ref{a1}-\ref{a4})
we obtain that $\eta^{gh}$ must vanish unless $g= h^{-1}$.
This means that our metric has only $N_G-1$ independent components,
which we shall denote as $E_g$:
\begin{equation}
\eta^{gh} \; = \; E_g \delta^{g(h^{-1})},
\;\;\; E_g \in \a,\; g \in G'.  \label{me3} \end{equation}

If we require that the constructed metric gives rise to a semi-norm,
we should restrict ourselves to such metrics, which are positive definite.
For the algebra of $\c$-valued functions this is equivalent to the
choice of real, non-negative $E_g$. The last issue, which we want to
point out is the question of degeneracy. We say that the metric is
non-degenerate if the condition $\eta(a^\star,a)=0$ implies that $a=0$.

The question, which we would like to raise next, is whether
this formal construction of the metric can be translated
to the metric properties of our base space, i.e., the group $G$.
It is important that in the construction of physical theories we can
have the picture of the underlaying base space rather than only of
the algebra $\a$. Therefore, we would like to have the
possibility of introducing both the distances and the concept
of the nearest neighbors.

We use the following definition for the distance $d(p,q)$ between two
points $p$ and $q$ of the base space \cite{CON3},
\begin{equation}
d(p,q) \; = \; \sup_{\eta(da,da^*) \leq 1}
 |p(a) - q(a)|.
\label{me4} \end{equation}
We have identified the base space as the space of
characters on the algebra $\a$, so that the definition
(\ref{me4}) makes sense for arbitrary $\a$. In our case,
of course, $p(a) \equiv a(p)$. The inequality
$\eta(da,da^\star) \leq 1$ means that the function
on its left-hand side
is majorized by the constant function $1$.

Before we present a few simple examples let us observe some general
properties of the metric. First, the metric does not have
to be symmetric, i.e., $\eta(u,v) \not= \eta(v,u)$.
However, after integrating out the result using the Haar
integration on $G$ we recover the symmetry.

The distances are, by definition, positive numbers from
the field $\f$, so in our case, where $\f=\c$,
they are real positive numbers.
Of course, the definition (\ref{me4}) implies the triangle
inequality:
\begin{equation}
d(p,q) \; \leq \; d(p,r) +  d(r,q),
\label{me5} \end{equation}
for any $p,q,r$.

Finally, let us introduce the notion of the
nearest neighbors of a point $h$, which shall
be all such elements of the base space of the form
$h\odot g, h\odot (g^{-1})$,
where $g \in G'$ and $E_g \not= 0$.

Now, let us proceed with the examples.
\begin{itemize}
\item {\sl $\z$ with a trivial metric}

Let us take the functions $E_g$ determining the metric $\eta$ to be
zero for $g \not=1$ and $E_1=1$. Then, the condition
$\eta(da,da^*) \leq 1$ simplifies
to $(a(p) - a(p+1))^2 \leq 1$, and one easily finds the distance
between $n,m \in \z$:
$$     d(n,m) \; = \; | n - m |.  $$
We can now draw a picture representing this base space.
If we connect the nearest neighbors with a link, then each
element has two nearest neighbors at the distance $1$ and
we obtain the image shown in Fig.1, which is the most natural
representation for $\z$.

\item {\sl $\z$ with a non-trivial metric}

Let us take $E_1=1$ and $E_2=1$ with all other $E_n$ vanishing.
The condition $\eta(da,da^*) \leq 1$ takes now the form:
$$ \left( (a(n)-a(n+1) \right)^2 + \left( (a(n)-a(n+2) \right)^2 \;
\leq \; 1,$$
and we see that the distances are different from those in the previous
example. The general formula is rather complicated, we shall only
mention that $d(n,n+1)=d(n,n+2)=1$,
$d(n,n+3)=\frac{1}{\sqrt{2}}+1$ and for
large $m$ the distance $d(n,n+m)$ is proportional
to $\frac{1}{\sqrt{5}}m$.
This result is presented in the picture Fig.2, where
we see that each point has now four nearest neighbors.

\quad We can go on further with more sophisticated choices of the metric
$\eta^{gh}$, deriving in each case the corresponding pictorial
representation. Of course, we do not have to deal with infinite
groups, one may as well take $\z_N$, in such case the resulting
diagram will be similar, though, of course, it would have
a topology of a circle.

\item {\sl $S_3$ with a non-trivial metric}

As the last interesting example we produce the $S_3$ group with
a rather complicated type of metric. Let $a$ and $b$ be the two
generators of $S_3$, such that $a^2=b^2=(ab)^3=\hbox{id}$. We take
$E_a=E_b=E_{aba}=1$ and that all other coefficients of the metric
vanish. Now, we have three nearest neighbors for each point
of $S_3$. The precise values of distances are rather difficult
to calculate and we shall not give these values here. What
interests us more is the picture we get by connecting
all elements with their nearest neighbors. The object we
obtain is presented in Fig.3. We easily
recognize that its topology (if we look at
the rectangular walls) is that of the Moebius strip.
This illustrates that the metric on the finite dimensional
objects may, in some sense, generate 'nontrivial topology'
of the resulting lattices.
\end{itemize}

\section{FIELD THEORIES}
\def\u{{\cal U}}
\def\hh{{\cal H}}

Having the metric and the formalism of the differential
geometry, we can construct field theories
for such spaces. The general procedure and some examples
of unitary gauge theories were presented in our
earlier work \cite{JA}, here we want to concentrate on
different aspects of the theory.

Our basic algebra $\a$ is again the commutative algebra of
complex valued functions on the group $G$. The unitary group
of this algebra, $\u(\a)$, contains all functions valued
in the circle $S^1$. The algebra $\a$ is generated by
$\u(\a)$ or any of its subgroups.

\subsection{Discretized Target Space}

To build a physical theory one requires a hermitian
vector bundle over the base space or, equivalently, a
projective module over $\a$. Taking a hermitian module we may
construct the simplest action in the usual way:
\begin{equation}
{\cal S} \; = \; \int < d m | d m > + V(<m|m>),
\label{r1} \end{equation}
where $m$ are the elements of the module and $V$ is an
arbitrary potential function. This approach has been
dealt with in many works \cite{CON}-\cite{CGV}.

However, this is not what we seek now, as this would not lead
us to theories having a discretized target space. The desired
formalism seems to be similar to this of the sigma models where
we take the group valued fields. If we take $U$ to be an element
of any group generating the algebra $\a$, then the
proposed expression for the action,
\begin{equation}
{\cal S} \; = \; \int \frac{1}{2} \eta \left( dU^\star, dU \right)
+ \tilde{V}(U),
\label{r2} \end{equation}
makes perfect sense. The action (\ref{r2}) is quite natural,
it contains both the 'kinetic' and the 'potential' terms, the latter
must be however restricted, so that the value of ${\cal S}$
is real. From the point of view of field theory the 'kinetic'
term describes the dynamics of the field and the other one
its self-interaction. However, we shall see later another, more
intuitive, interpretation.

We shall use this prescription to construct simple
models of the discrete geometry. We take the group $\hh$ to
be any subgroup of $\u(\a)$ and the action precisely as
defined in (\ref{r2}), with the integration on the algebra
being the already introduced Haar integration.
Taking into account the form of the metric
(\ref{me3}) and the rules of the differential calculus
(\ref{a1}-\ref{a8}) we may rewrite the action as follows:
\begin{equation}
{\cal S} \; = \; \int_G \; \left( - \frac{1}{2}
\sum_{g \in G'} E_g (U^\star - R_g U^\star) ( U - R_g U) \right) + V(U).
\label{s2a}
\end{equation}
Using the properties of the integration we finally obtain:
\begin{equation}
{\cal S} \; = \; \int_G \; \left( - \sum_{g \in G'} E_g \Re
\left( U^\star ( R_g U) \right) \right) + V(U),
\label{s2b} \end{equation}
where we have omitted the constants coming from $U^\star U$ terms.
Now, we shall attempt to rewrite (\ref{s2b}) in a slightly more
convenient and recognizable form. Remember that the Haar integration
is nothing else but the sum of over all elements of the group $G$:
\begin{equation}
\int_g \; f \; = \; \sum_{h \in G} f(h),
\label{s2d} \end{equation}
so that the potential term splits into the sum of
independent contributions from each point of the base space:
\begin{equation}
\int_G V \; = \; \sum_{h \in G} V\left( U(h) \right).
\label{s2e} \end{equation}
The 'kinetic' term is more interesting. Remember that the
coefficients $E_g$ define the metric structure of the group $G$.
Having introduced the natural idea of nearest neighbors
we may see that the sum over $g \in G'$, which appears in the
definition, combined with the Haar integration is nothing
else but the sum over nearest neighbors with certain
weights. Therefore we can rewrite this term as:
\begin{equation}
\int_G \; \sum_g E_g U (R_g U) \; = \!\!\!\!\!\!\!
\sum_{\stackrel{h,g \in G}{\hbox{\tiny nearest neighbors}}}
\!\!\!\!\!\!\!
W(g,h) U(h) U(g),
\label{s2f} \end{equation}
where $W(g,h)$ is the weight, which equals $E_{(h^{-1}g)}(h)$.

\subsection{Examples}

Having constructed the general form of the action we can now
present a few interesting examples. We restrict
ourselves only to the most spectacular situations
as we want only to demonstrate the analogies between
the models of statistical physics and of noncommutative
geometry.

\begin{itemize}
\item {\sl The Ising Model}

Let us take the group $\hh$ to be the group
of $\z_2$ valued functions on $\z$. Because for any $\z_2$ valued
function $U^2=1$, the potential term can be reduced to the linear form
$V(U) \sim U$. If we fix the metric to be the standard metric on $\z$,
as in the first example of the previous section, we get the following action:
\begin{equation}
{\cal S} \; = \; \sum_{n \in \ze} \alpha U(n) U(n+1) + \gamma U(n),
\label{s2h} \end{equation}
where $\alpha,\gamma$ are arbitrary real constants. We easily
recognize that the action describes precisely the Ising model.
Note that the 'kinetic' term of our field theory has now the meaning
of the interaction between the nearest neighbors,
while the 'potential' term has no other specific interpretation
apart from being the interaction with some external fields.
The constant $\alpha$ sets the value of the
gap between the energy levels of the model. The path integral
is now simply the partition function of the Ising model.

If the potential term is absent the action ${\cal S}$ possesses
a global symmetry as the change $U \to -U$ leaves the action
invariant.

\item {\sl The Ising Model with a Non-standard Metric}

As the next example we take again the same group and the
same base space but with a different metric. This time we
assume that the metric is as in the second example of the
last section, i.e. $E_1=E_2=1$ and that all other coefficients
vanish. Then, after similar steps as in the
previous case, we obtain the following action:
\begin{equation}
{\cal S} \; = \; \sum_{n \in \ze} \alpha \left( U(n) U(n+1) + U(n)
U(n+2) \right) + \gamma U.
\label{s2i} \end{equation}

This again is a variation of the Ising model, however, on
a slightly modified lattice with each point having
four neighbors, as symbolically represented in Fig.2.

\item {\sl The three-state Potts model}

Consider now the group of $\z_3$ valued functions.
Following the same procedure as in two previous cases we
construct the action, taking as the metric over $\z$
again the standard metric (the same as in the first example).
Then the interaction term reads:
\begin{equation}
{\cal S} \; = \; - \sum_n \alpha \Re \left( U(n)^\star U(n+1) \right).
\label{s2j} \end{equation}
This action (modified slightly by adding an appropriate potential
term) can be recognized as the one describing a three-state
Potts model. By changing the metric we may, of course,
modify the interaction by increasing the
number of the nearest neighbors.
\end{itemize}

All these examples deal with one-dimensional models but the
generalization to higher dimensions is straightforward.
For instance, one has to take the group $\z^n$ to obtain
the $n$-dimensional generalization of considered models.
If we want to restrict theories to a finite base space
(so that the action is a finite sum) we take the base space to
be $\z_N$, and take the limit $N \to \infty$ to recover the
case of the $\z$-based model.

These examples illustrate that the simple models of
statistical physics have their interpretation as a field
theory constructed in the framework of noncommutative geometry.
They are all built in a rather simple fashion, using the commutative
group $\z$ as the base space and a finite commutative unitary
group as the target space. One may, of course, attempt to go beyond
that and use the same tools to construct more sophisticated theories,
for the nonabelian groups, for instance.
Another new possibility is to construct gauge theories,
extending the observed global symmetries to the
local ones. We shall see the exemplary construction
in the next section.
\section{THE GAUGE THEORY}

Now, we shall briefly outline the prospects of creating
the gauge theory by exploiting the symmetry that we have
noticed in the last section. The natural extension
of the observed global symmetry is the group
$\hh$ itself, so we propose it as a gauge symmetry group.

Following the construction procedures from our earlier work
\cite{JA} we take the gauge connection one-form $\Phi$:
\begin{equation}
\Phi \; = \; \sum_{g \in G'} \Phi_g \chi^g,
\label{e1} \end{equation}
where the coefficients $\Phi_g$ belong
to the algebra $\a$. It would be convenient to use the shifted connection,
$\Psi_g = 1 - \Phi_g$, as then all the expressions simplify considerably.
Since the group is unitary we have the hermicity constraint, which is:
\begin{equation}
\Psi_g^* \; = \; R_g ( \Psi_{(g^{-1})}).
\label{e2} \end{equation}
The curvature two-form $F = d\Phi + \Phi \Phi$ expressed in terms
of $\Psi$ reads:
\begin{equation}
F_{gh} \; = \; \Psi_g R_g( \Psi_h ) - \Psi_{(h \odot g)}
\label{e3} \end{equation}
where we identify $\Psi_e$ with $1$. Having the metric
$\eta$ of the form (\ref{me3}) we can construct all possible
Yang-Mills type actions:
\begin{eqnarray}
S_1 & = & \int_G \sum_{g,h} E_g E_{(h^-1)} \left( \Psi_g
\Psi_g^\star - 1 \right) \left( \Psi_h^\star \Psi_h - 1 \right),
\label{e3a} \\
S_2 & = & \int_G \sum_{g,h} E_g E_{(h^{-1})} \left( \Psi_g
R_g \Psi_h - \Psi_{(h \odot g)} \right) \left( \Psi_g
R_g \Psi_h - \Psi_{(h \odot g)} \right)^\star,
\label{e3b} \\
S_3 & = & \int_G \sum_{g,h} E_g E_{h} \left( \Psi_g
R_g \Psi_h - \Psi_{(h \odot g)} \right) \left( \Psi_h
R_h \Psi_g - \Psi_{(g \odot h)} \right)^\star,
\label{e3c}
\end{eqnarray}

We shall concentrate now on the particular case of the Ising model.
The metric is defined by taking $E_1=1$ and $E_g=0$ for $g \not= 1$.
This fixes the actions (\ref{e3a}-\ref{e3c}) to take
the following form:
\begin{eqnarray}
S_1 & = & \sum_n \left( \Psi_1(n) \Psi_1^\star(n) - 1 \right)^2,
\label{e4a} \\
S_2 & = & S_1,
\label{e4b} \\
S_3 & = & \sum_n |\left( \Psi_1(n) \Psi_1(n+1) - \Psi_2(n)
\right)|^2.
\label{e4c}
\end{eqnarray}
First, let us notice that only the fields $\Psi_1$ and $\Psi_2$
contribute to the action, which follows from our choice of the metric.
Moreover, this choice makes the action $S_1$ to have no
interaction terms, which causes that for every $n$ the value of
$\Psi_1(n)$ is independent of other values of this field.
The situation is somehow different in the third possible action,
where we have both the interaction term for $\Psi_1$ and the
interaction between $\Psi_1$ and $\Psi_2$.

The model described by the first action (\ref{e4a}) is not
interesting from the physical point of view, as it describes
a completely non-interacting system. We shall not discuss here
the other action and its properties, as it does not
resemble any model of statistical physics.

\section{CONCLUSIONS}

We have shown a way to construction a class
of field theories in the discrete geometry,
which have their target space discretized.
We found that some of them correspond exactly
to the well-known models
of statistical physics. We were able to modify them
slightly by changing the free parameters of our
construction, which were the metric and the potential.

Let summarize the most important facts about
the construction.
The space of fields was chosen to
be a subgroup of the unitary group of the algebra $\a$,
which determined the target space. The form of the
interaction was dependent only on the metric of the base
space and it appeared in the action as a 'kinetic' term.
We also allowed a potential term. This determined the action
and the model completely.

We believe that the correspondence with the field theory, which
we presented in this paper for the Ising model and the three-state
Potts model, can be extended to many other systems. Moreover,
using this method we may be able to analyze and
compare their properties from another angle, we may also use
the methods to create other models, by fitting the algebra $\a$,
the subgroup $\hh$ and the metric $\eta$. Whether such models
would exhibit any interesting features
remains an open question.

Finally, we presented a method of building the gauge theory, using
a discretized group of gauge symmetries. It seems, however, that
the resulting models, at least in the studied case,
were of little physical meaning.

In our considerations we used the commutative algebra $\a$ of
complex valued functions on $G$. Let us mention here that
the same analysis may be repeated for algebras over $\z$.
In such case the algebra $\a$ would be simply defined as generated
by the group $\hh$. Let us point out that in such
case one does not have to restrict oneself to the abelian
groups. Such situation would be probably the most interesting one.

\def\v#1{{\bf #1}}

\newpage
\null \vskip-1truein
\unitlength=1cm
\begin{picture}(12,2)
\put(1.5,1){\line(1,0){10}}
\put(2,1.3){\makebox(0,0){\tiny -5}}
\put(3,1.3){\makebox(0,0){\tiny -4}}
\put(4,1.3){\makebox(0,0){\tiny -3}}
\put(5,1.3){\makebox(0,0){\tiny -2}}
\put(6,1.3){\makebox(0,0){\tiny -1}}
\put(7,1.3){\makebox(0,0){\tiny 0}}
\put(8,1.3){\makebox(0,0){\tiny 1}}
\put(9,1.3){\makebox(0,0){\tiny 2}}
\put(10,1.3){\makebox(0,0){\tiny 3}}
\put(11,1.3){\makebox(0,0){\tiny 4}}
\multiput(2,1)(1,0){10}{\circle*{0.2}}
\put(6.5,0.4){\makebox(0,0){Fig.1 $\z$ with trivial metric}}
\end{picture}

\vspace{1cm}

\unitlength=1cm
\begin{picture}(12,4)
\put(1.5,3){\line(1,0){10}}
\put(1,2.25){\line(1,0){11}}
\put(2,3.3){\makebox(0,0){\tiny -9}}
\put(3,3.3){\makebox(0,0){\tiny -7}}
\put(4,3.3){\makebox(0,0){\tiny -5}}
\put(5,3.3){\makebox(0,0){\tiny -3}}
\put(6,3.3){\makebox(0,0){\tiny -1}}
\put(7,3.3){\makebox(0,0){\tiny 1}}
\put(8,3.3){\makebox(0,0){\tiny 3}}
\put(9,3.3){\makebox(0,0){\tiny 5}}
\put(10,3.3){\makebox(0,0){\tiny 7}}
\put(11,3.3){\makebox(0,0){\tiny 9}}
\put(1.5,2){\makebox(0,0){\tiny -10}}
\put(2.5,2){\makebox(0,0){\tiny -8}}
\put(3.5,2){\makebox(0,0){\tiny -6}}
\put(4.5,2){\makebox(0,0){\tiny -4}}
\put(5.5,2){\makebox(0,0){\tiny -2}}
\put(6.5,2){\makebox(0,0){\tiny 0}}
\put(7.5,2){\makebox(0,0){\tiny 2}}
\put(8.5,2){\makebox(0,0){\tiny 4}}
\put(9.5,2){\makebox(0,0){\tiny 6}}
\put(10.5,2){\makebox(0,0){\tiny 8}}
\put(11.5,2){\makebox(0,0){\tiny 10}}
\multiput(2,3)(1,0){10}{\circle*{0.2}}
\multiput(1.5,2.23)(1,0){11}{\circle*{0.2}}
\multiput(2,3)(1,0){10}{\line(2,-3){0.5}}
\multiput(2,3)(1,0){10}{\line(-2,-3){0.5}}
\put(6.5,1.4){\makebox(0,0){Fig.2 $\z$ with non-trivial metric}}
\end{picture}

\vspace{1cm}

\unitlength=10mm
\begin{picture}(12,4)
\put(4,4){\circle*{0.2}}
\put(4,2.5){\circle*{0.2}}
\put(9,4){\circle*{0.2}}
\put(9,2.5){\circle*{0.2}}
\put(5.5,1){\circle*{0.2}}
\put(7.5,1){\circle*{0.2}}
\put(3.7,4){\makebox(0,0){\tiny id}}
\put(3.7,2.5){\makebox(0,0){\tiny a}}
\put(9.3,4){\makebox(0,0){\tiny b}}
\put(9.3,2.5){\makebox(0,0){\tiny ba}}
\put(5.5,0.6){\makebox(0,0){\tiny ab}}
\put(7.5,0.6){\makebox(0,0){\tiny aba}}
\put(4,4){\line(1,0){5}}
\put(4,2.5){\line(1,0){5}}
\put(4,2.5){\line(0,1){1.5}}
\put(9,2.5){\line(0,1){1.5}}
\put(5.5,1){\line(1,0){2}}
\emline{5.5}{1}{1}{4}{2.5}{2}
\emline{7.5}{1}{3}{9}{2.5}{4}
\emline{5.5}{1}{5}{7.25}{2.5}{6}
\emline{6.5}{1.857}{7}{5.75}{2.5}{8}
\multiput(5.75,2.5)(-0.175,0.15){10}{\circle*{0.01}}
\multiput(7.5,1)(-0.175,0.15){5}{\circle*{0.01}}
\multiput(7.25,2.5)(0.175,0.15){10}{\circle*{0.01}}
\put(6.5,-0.5){\makebox(0,0){Fig.3 $S_3$ with non-standard mectric}}
\end{picture}
\vfill

\begin{thebibliography}{10}

\bibitem{CON} {A.Connes,
{\em Non-commutative differential geometry, de Rham homology and
non-commutative algebra},
Publ. Math. IHES \v{62} (1986), 44-144}

\bibitem{CON1} {A.Connes,J.Lott,
{\em  Particle models and non-commutative geometry}, \\
Nucl.Phys. (Proc. Suppl.) \v{B18},
(1990) 29-47}

\bibitem{CON2} {A.Connes,
{\em Essays on physics and non-commutative geometry},
in: "The Interface of Mathematics
and Particle Physics", eds D.Quillen, G.Segal and S.Tsou, Oxford
University Press (1990)}

\bibitem{CON3} {A.Connes,
{\em Geometrie non Commutative}, Intereditions, Paris (1990) }

\bibitem{CON4} {A.Connes,
{\em Lectures at the Les Houches Summer School 1992},}

\bibitem{Coq}
{R.Coquereaux,
{\em Non-commutative geometry and theoretical physics},
Journal of Geometry and Physics, 1990}

\bibitem{CGV}
{R.Coquereaux,G. Esposito-Far\`ese, G.Vaillant,
{\em Higgs field as Yang-Mills field and discrete symmetries},
Nucl. Phys. \v{B353}, (1991) 689-706}

\bibitem{JA}
{ A.Sitarz
{\em Noncommutative Geometry and Gauge Theory on Discrete
Groups}, preprint TPJU-7/92, hep-th {\#}9210098}

\bibitem{HUANG}
{Kerson Huang,
{\em Statistical Mechanics, New York 1963} }
\end{thebibliography}
\end{document}